\documentclass[preprint,showpacs,preprintnumbers,amsmath,amssymb]{revtex4}

\usepackage[dvips]{epsfig}
\usepackage{color}

\begin{document}

\title{Dilepton distributions at backward rapidities}

\author{M.A. Betemps$^{1,2}$}
\email{marcos.betemps@ufrgs.br} 
\author{M.B. Gay Ducati$^1$}
\email{beatriz.gay@ufrgs.br}  
\author{E.G. de Oliveira$^1$}
\email{emmanuel.deoliveira@ufrgs.br}

\affiliation{$^{1}$ High Energy Physics Phenomenology Group (GFPAE),
Instituto de F\'{\i}sica, Universidade Federal do Rio Grande do Sul,
Caixa Postal 15051, CEP 91501-970, Porto Alegre, RS, Brazil\\
$^{2}$ Conjunto Agrot\'ecnico Visconde da Gra\c ca, CAVG,\\
Universidade Federal de Pelotas,
Caixa Postal 460, CEP 96060-290, Pelotas, RS, Brazil
}

\date{\today}

\begin{abstract}
The dilepton production at backward rapidities in $pAu$ and $pp$
collisions at RHIC and LHC energies is investigated in the dipole
approach. The results are shown through the nuclear modification ratio
$R_{pA}$ considering transverse momentum and rapidity spectra. The
dilepton modification ratio presents interesting behavior at the
backward rapidities when compared with the already known forward ones,
since it is related with the large $x$ kinematical region that is
being probed.  The rapidity dependence of the nuclear modification
ratio in the dilepton production is strongly dependent on the Bjorken
$x$ behavior of the nuclear structure function ratio
$R_{F_{2}}=F_{2}^{A}/F_{2}^{p}$. The $R_{pA}$ transverse momentum
dependence at backward rapidities is modified due to the large $x$
nuclear effects: at RHIC energies, for instance, the ratio $R_{pA}$ is
reduced as $p_T$ increases, presenting an opposite behavior when
compared with the forward one. It implies that the dilepton production
at backward rapidities should carry information of the nuclear effects
at large Bjorken $x$, as well as that it is useful to investigate the
$p_T$ dependence of the observables in this kinematical regime.
\end{abstract}

\pacs{27.75.-q, 13.85.Qk}
\maketitle

\section{Introduction}

The hadron transverse momentum spectrum measured by the RHIC
experiments \cite{RHICresults} is one of the most accessible
distributions carrying information about the high density nuclear
environment. At forward rapidities, the transverse momentum
distribution for hadrons presents results compatible with the Color
Glass Condensate (CGC) description of the saturated regime at high
energies \cite{RHICCGC}. These results are investigated through a
nuclear modification ratio, which relates $dAu$ and $pp$ collisions.
It is well known that the transverse momentum dependence of this ratio
shows a pronounced Cronin peak at central rapidity, which is
suppressed at forward ones. For central rapidities the Cronin peak is
due to the multiple scatterings of the projectile constituents with
the dense target. At large rapidities the Cronin peak suppression is
understood as a signal of the saturation phenomena intrinsic of the
Color Glass Condensate (CGC). However, at backward rapidities, the
recently measured hadron nuclear modification ratio shows a pronounced
peak \cite{backphenix}, which should be due to final states
interactions.

In this work the dilepton production at high energies in the backward
rapidity region is investigated in the proton-nucleus and
proton-proton collisions. The analysis at forward rapidities was
already done \cite{PRD70116005} and the results show that the dilepton
is also a suitable observable to investigate the Color Glass
Condensate. Here, the interest to explore the backward region in the
dilepton sector relies on the study of the nuclear effects at large
and small $x$ and on the comparison with the Cronin effect in this
region.  In this kinematical region, the nucleus interacts by means of
large Bjorken $x$ partons, and no saturation effects are expected
consequently.  In this regime, the proton interacts through the small
$x$ partons.

It has been shown \cite{PRD70116005} that the dilepton production at
forward rapidities, analyzed in the context of the Color Glass
Condensate in $pAu$ collisions, presents the same features of the
Cronin effect at forward rapidities, implying that this should be
considered as an initial state effect. However, at backward
rapidities, the Cronin peak present in the hadrons RHIC data
\cite{backphenix} is still not understood, and could be related to
final state effects.
The dileptons analyzed in this kinematical region, since they do not
interact strongly with the final environment, may not present the same
behavior of the hadron spectra. Although it could imply the dilepton
production to be symmetric when forward-backward regions are compared,
in contrast to the different behavior of the hadron spectra in both
regions, we verify that the dilepton production at backward rapidities
is strongly modified due to large $x$ nuclear effects.

This work is organized as follows. In the next section the dipole
approach is presented for the dilepton production at backward
rapidities.  In the Sec.\ 3 the nuclear partonic parametrizations
employed in this work as well as a comparison between then are
presented. In the Sec.\ 4 the results are discussed. Our conclusions
are left to the last section of this work.

\section{Dilepton Production at Backward Rapidities}

To evaluate the dilepton production at backward rapidities, an
adequate treatment of the approach employed for the forward ones needs
to be considered. The investigation at forward rapidities in $pA$
collisions requires to describe the nucleus as a high density
saturated system (CGC), and the dileptons are produced by the decay of
a virtual photon emitted by the quark from the proton, which interacts
with the nuclear dense medium (basically a dipole approach in the
momentum space with a specific configuration of the nucleus). In this
case the dilepton cross section can be factorized considering the wave
function of the fluctuation $q\gamma^*$ and the interaction cross
section of a quark with the nucleus, which takes into account the high
density effects. However, at backward rapidities the nucleus interacts
by means of large Bjorken $x$ partons, being no more described by a
high density QCD system.  The proton interacts through very small $x$
partons implying that the dipole picture could be applied in this
kinematical regime.

\begin{figure}[h]
  \centerline{\scalebox{0.85}{\includegraphics*[155pt,615pt][300pt,740pt]{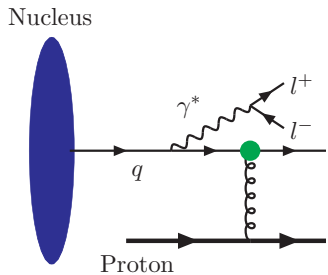}}}
    \center{%
	{\caption{
      \label{dybrem}
      In the dipole approach,
      dilepton production looks like a bremsstrahlung. A quark 
      or an anti-quark from the
      projectile hadron (nucleus) scatters off the target color field  and radiates a
      photon ($\gamma^*$) with mass $M$ (before or after the quark scatters), 
which subsequently decays into the lepton pair. }  
    }  }
\end{figure}

In the dipole approach at backward rapidities, the quark of the
incident hadron (nucleus) fluctuates into a state containing a quark
and a massive photon. The interaction with the target (proton)
provides that the virtual photon is freed, decaying into a lepton pair
\cite{NuclRauf} (one of the diagrams is shown in the Fig.\
\ref{dybrem}).
 
We are interested in the study of the dilepton production in a region
where the interaction time of the projectile quark with the proton
target is much shorter than the $q\gamma^*$ fluctuation time, meaning
large coherence length.  In order to apply the dipole picture at
backward rapidities, we need to exchange the nucleus with the proton
when comparing with the dipole picture at forward rapidities. In doing
such procedure, the $q\gamma^*$ fluctuation time will be larger than
the interaction time, since the coherence length $l_c$ has the
dependence $l_c\propto 1/x_1$ in the case studied here. Indeed, the
coherence length is important in the study of nuclear targets, where
it is a fundamental quantity controlling nuclear effects
\cite{nuclbroad}.  For the nuclear targets large coherence length
needs to be reached if shadowing effect are under investigation, which
implies that the approach should be applied only to smaller Bjorken
$x$ \cite{Armesto:2002ny}.  Another interesting point that arises here
is that the $p_T$ broadening due to the nuclear multiple scattering
verified at forward rapidities \cite{nuclbroad} should not be present
at backward rapidities, since the nuclear projectile will interact
with a proton target and will not probe a large and very dense system.

The advantage of this formalism is that the dilepton cross section can
be written in terms of the same color dipole cross section as
small-$x$ Deep Inelastic Scattering (DIS).  Although diagrammatically
no dipole is present in bremsstrahlung, the dipole cross section
arises from the interference of the two bremsstrahlung diagrams, as
shown in Ref.~\cite{PRDrauf} in a detailed derivation.

The cross section for the radiation of a virtual photon from a quark
(with momentum fraction $x_2$) of the nucleus scattering off a proton
at high density at backward region can be written in a factorized form
as \cite{PRD67114008,DYdipole2},
\begin{eqnarray}
\frac{d\sigma^{DY_{back}}}{dM^{2}dyd^{2}p_{T}}\!\!=\!\!\frac{\alpha_{em}^{2}}{6\pi^{3}M^{2}}\int_{0}^{\infty}\!\!\!d\rho W(x_{2},\rho,p_{T})\sigma_{dip}(x_{1},\rho),
\end{eqnarray}
where $p_{T}$ is the dilepton transverse momentum, $M$ is the dilepton
mass, $y$ the rapidity, $\rho$ is the dipole transverse
separation. The variables $x_1$ and $x_2$ are defined in the usual way
$x_{1 \choose 2}=\sqrt{\frac{M^2 +p_T^2}{s}}e^{\pm y}$, with $s$ being
the squared center of mass energy. As can be seen from the definition
of $x_1$ and $x_2$, backward rapidities will provide large $x_2$ and
small $x_1$. We are investigating dilepton production in $pp$ and $pA$
collisions. In a symmetric collision, the lab frame is equivalent to
the center of mass frame considering colliders. In the case of
asymmetric collisions, e.g. $pA$, the center of mass moves
longitudinally in the lab frame. It provides that a largest interval
in Bjorken $x$ could be explored in the asymmetric collisions
\cite{asymmcoll} if the lab frame is under consideration. In this work
the ratio between $pA$ and $pp$ collisions will be investigated
considering the rapidity in the lab frame.

Based on the dipole approach, the function $W(x_2,\rho,p_{T})$ is
given by \cite{PRD67114008},
\begin{widetext}
\begin{eqnarray}\nonumber
W(x_2,\rho,p_{T}) & = & \int_{x_{2}}^{1}\frac{d\alpha}{\alpha^{2}}F_{2}^{A}(\frac{x_{2}}{\alpha},M^{2})\left\{ [m_{q}^{2}\alpha^{2}+2M^{2}(1-\alpha)^{2}]\left[\frac{1}{p_{T}^{2}+\eta^{2}}T_{1}(\rho)-\frac{1}{4\eta}T_{2}(\rho)\right]\right.\\
 & + & \left.[1+(1-\alpha)^{2}]\left[\frac{\eta p_{T}}{p_{T}^{2}+\eta^{2}}T_{3}(\rho)-\frac{1}{2}T_{1}(\rho)+\frac{\eta}{4}T_{2}(\rho)\right]\right\}.
\end{eqnarray}
\end{widetext}

In the above equation $\alpha$ is the momentum fraction of the quark carried by the
virtual photon, $\eta^2=(1-\alpha)M^2 +\alpha^2m_q^2$, $m_q$ is the
quark mass ($m_q=0.2$ GeV) and the functions $T_{i}$ are given by,
\begin{eqnarray*}
T_{1}(\rho) & = & \frac{\rho}{\alpha}J_{0}(\frac{p_{T}\rho}{\alpha})K_{0}(\frac{\eta\rho}{\alpha})\\
T_{2}(\rho) & = & \frac{\rho^{2}}{\alpha^{2}}J_{0}(\frac{p_{T}\rho}{\alpha})K_{1}(\frac{\eta\rho}{\alpha})\\
T_{3}(\rho) & = & \frac{\rho}{\alpha}J_{1}(\frac{p_{T}\rho}{\alpha})K_{1}(\frac{\eta\rho}{\alpha}).
\end{eqnarray*}

Here, a nuclear structure function
$F_{2}^{A}(\frac{x_{2}}{\alpha},M^{2})$ with a $x_{2}$ dependence is
considered in order to take into account the nuclear interaction. For
$pp$ collisions the nuclear structure function
$F_2^A(\frac{x_2}{\alpha}, M^2)$ needs to be replaced by the proton
structure function $F_2^p(\frac{x_2}{\alpha}, M^2)$. The dipole cross
section is evaluated with the argument $x_{1}$ and the same dipole
cross section extracted from the deep inelastic scattering HERA data
should be applied.

\begin{figure}[h]
\includegraphics[scale=0.4]{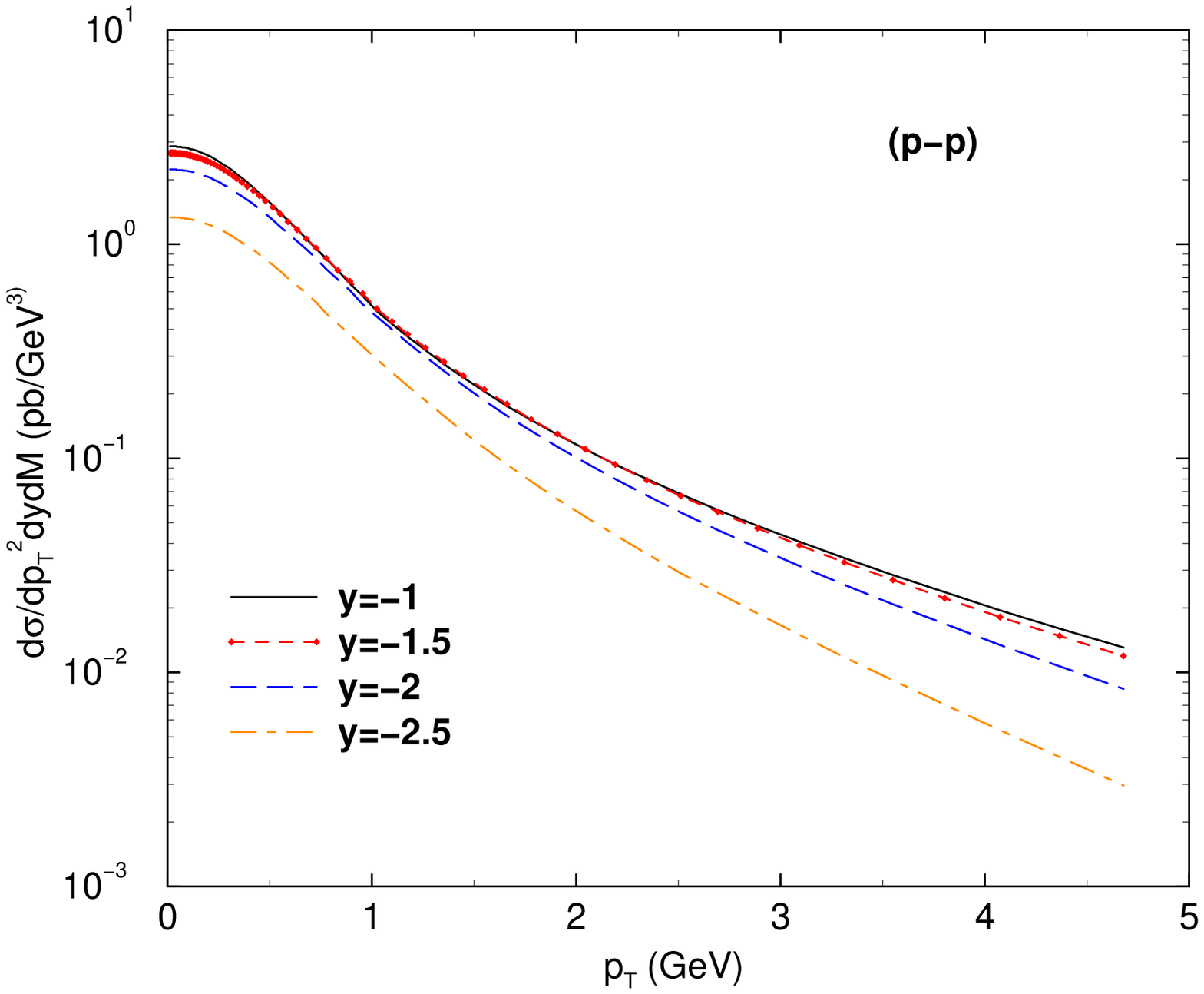}
\includegraphics[scale=0.4]{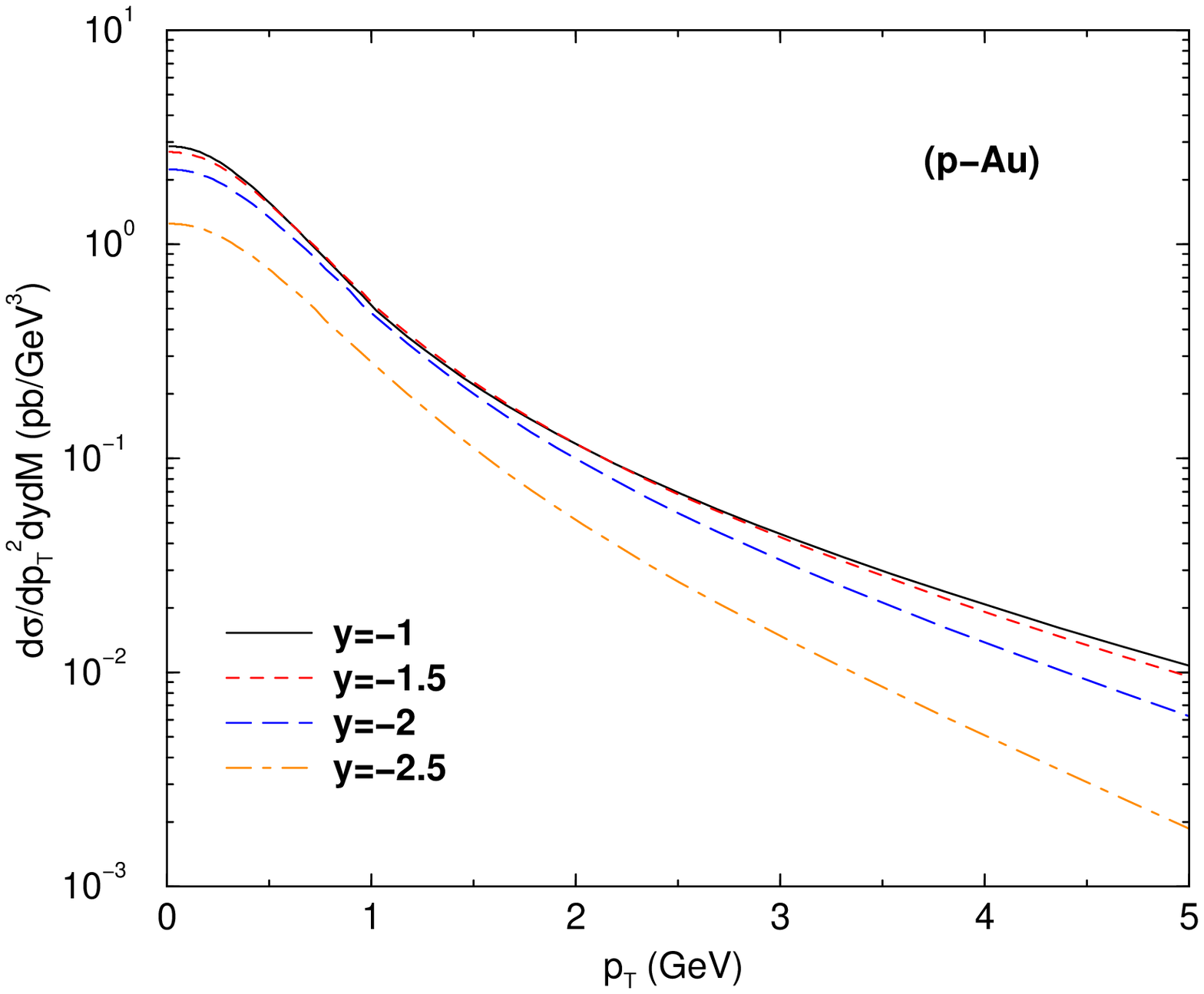}
\caption{\label{cap:Result-for-pp} Dilepton transverse momentum distribution
for $pp$ and $pA$ (Au -- considering EKS parametrization) collision at
RHIC energies for dilepton mass $M=6$ GeV.}
\end{figure}

The dipole cross section in the phenomenological model proposed by
Golec-Biernat and W\"usthoff \cite{GBW} (GBW), which has produced a
good description of HERA data in both inclusive and diffractive
processes, was employed here. It is constructed interpolating the
color transparency behavior $\sigma_{dip} \sim \rho^2$ for small
dipole sizes and a flat (saturated) behavior for large dipole sizes
$\sigma_{dip} \sim
\sigma_0$ (confinement). The expression has the eikonal-like form, 
\begin{eqnarray}
\sigma_{dip}(x_1,\rho)=\sigma_{0}\left[1-\exp\left(-\frac{\rho^{2}Q_{0}^{2}}
{4(x_1/x_{0})^\lambda}\right)\right], 
\end{eqnarray} 
where $Q^2_{0}=1$ GeV$^2$ and the three fitted parameters are
$\sigma_{0}=23.03$ mb, $x_{0}=3.04\times10^{-4}$ and $\lambda=0.288$. 

The $pp$ and $pAu$ transverse momentum distributions at RHIC energies
for backward rapidities can now be evaluated, considering a specific
parametrization to the nuclear structure function $F_{2}^{A}(x,M^{2})$
in the $pA$ case. Two distinct parametrizations for the nuclear
structure function will be used, which will be properly discussed in
the next section.  In this section EKS nuclear parametrization
\cite{EKS} was employed together with the GRV98 parton distribution
function \cite{GRV98} in order to obtain the $p_T$ distribution for
the $pA$ collision and the $F_2^p(x,M^2)$ is obtained from the GRV98
parametrization for the $pp$ collision. In the Fig.\
\ref{cap:Result-for-pp} the $pp$ and $pA$ transverse momentum
distributions for RHIC energies are shown and significative visible
differences between both distributions are not found. For this reason,
the ratio between $pA$ and $pp$ cross section should be useful to
investigate modifications in the nuclear cross section in comparison
with the $pp$ distribution. The ratio is defined in the following way
\begin{eqnarray}
R_{pA}=\left.\frac{d\sigma(pA)}{dp_{T}^{2}dydM}\right/A\frac{d\sigma(pp)}{dp_{T}^{2}dydM}.
\end{eqnarray}
The ratio $R_{pA}$ is normalized by $1/A$, since the nuclear
parton distributions are normalized by this factor.
The difference
between $pA$ and $pp$ calculations are due to the nuclear structure
function $F_{2}^{A}$, then any effect in the $p_{T}$ or rapidity
spectra should be related to the nuclear effects contained in this function.

The next-to-leading order calculation (NLO) in the collinear
factorization could be evaluated and compared with the results in the
dipole approach. However, the evaluation of the transverse momentum
distribution in the collinear factorization is not a straightforward
calculation and does not describe the normalization of the $p_T$
Drell-Yan dileptons \cite{resumma,gfai}. A resummation of large
logarithms in $p_T/M$ \cite{resumma,gfai} or the introduction of an
intrinsic transverse momentum \cite{intrisic} should be necessary to
avoid the divergence of the differential cross section at
$p_T<<M$. New approaches have been suggested in order to describe the
$p_T$ spectra of the dilepton (for a good discussion see
Ref. \cite{linnyk}), however, there is a strong dependence on
phenomenological parameters. Considering, for instance, the intrinsic
transverse momentum, the $p_T$ distribution is dependent on a
phenomenological parameter (see Ref.~\cite{PRDrauf} for a comparison
between the NLO calculations and the dipole approach, and
Ref.~\cite{linnyk} for a new approach). Of course, the ratio $R_{pA}$
should be finite since the $p_T$ divergence will be canceled, however,
a more reliable result needs to consider finite spectra. This
indicates that the dipole approach should be more adequated to
investigate the low $p_T$ dileptons at high energies, since there are
no free parameters and the cross section is finite at low $p_T$.

For sake of completeness, it should be stressed that some limitations
of the dipole approach at backward rapidities are in order.  Here, the
nucleus is considered by means of an integrated gluon distribution,
however, at LHC energies and more central rapidities, the Bjorken
$x_2$ (nucleus) reaches values around 0.002, where the consideration
of an integrated parton distribution could be questionable. At this
high energy limit, the effects of finite transverse momentum of the
incoming partons become important. A more robust treatment would need
to include the transverse momentum of the partons from the nucleus in
the initial state of the interaction.  This should be done considering
the $k_T$ factorization approach \cite{ciafaellisglr}, where the
off-shell partonic cross sections are convoluted with $k_T$
unintegrated parton densities $f_a(x,k_T^2,\mu^2)$. Considering
dilepton production, the $k_T$ factorization is investigated in the
Ref.~\cite{linnyk2} and compared with a phenomenological intrinsic
$k_T$ approach ($k_T$ factorization with on-shell partons), in order
to describe the transverse momentum distribution of the Drell-Yan
dilepton production. In spite of a reasonable data description, the
$k_T$ factorization overestimates the data and the intrinsic $k_T$
approach depends on phenomenological parameters (two parameters).

For the reasons presented above, we have focoused our analysis at
backward rapidities, and not at more central ones. The use of the
dipole approach together with the collinear factorization is justified
as the goal of this work is to investigate the nuclear effects in the
nuclear modification ratio $R_{pA}$ for dileptons, using nuclear
parametrizations to describe the $F_2^A$.  In the next section the
parametrizations for the nuclear structure function will be discussed
and compared.

\section{Nuclear Parton Distribution Function}

In this work the parametrizations for the nuclear parton distribution
function (nPDF) proposed by Eskola, Kolhinen and Salgado (EKS
parametrization) \cite{EKS} and by D. de Florian and R. Sassot (nDS
parametrization) \cite{nDS} are considered in order to describe the
nuclear effects in the $F_{2}^{A}$. In this section the main features
of each parametrization will be discussed.

Both nuclear parton distribution functions provide a global fitting to
fixed target experimental results, and consider the DGLAP equations
for the $Q^{2}$ evolution. The initial conditions are adjusted to
describe the DIS in lepton-nucleus collisions and the dilepton
production in proton-nucleus collisions.  Both nPDFs are used to
obtain an appropriated parametrization for the ratio
$R_{F_{2}}^{A}(x,Q_{0}^{2})= F_{2}^{A} / A F_{2}^{p}$, with
\begin{equation}
F_{2}^{A,p} = \sum_q e^2_q[x f_q^{A,p}(x,Q_0^2)+x f_{\bar q}^{A,p}(x,Q_0^2)],
\end{equation}
in which $f_q^{p}(x,Q_0^2)$ are the free proton parton distributions and $f_q^{A}(x,Q_0^2)$ are the nuclear distributions of parton flavor $q$. The parametrization for the parton distributions are constrained
by charge, baryon number, and momentum conservation. 

However, the method used by EKS and nDS differs with respect to their
approaches to parametrize the nPDFs. In the EKS approach, the nuclear
parton distributions are the free proton ones times a correction
factor: $f_q^{A}(x,Q_0^2)=R_q^A(x, Q_0^2)f_q^{p}(x,Q_0^2)$, and all
nuclear effects are enclosed in $R_q^A(x, Q_0^2)$. One consequence of
this definition is that the nPDFs are null for $x>1$, although they
should be non-zero for $x < A$. On the other hand, nDS uses a
convolution to relate nPDFs to the free proton PDFs:
\begin{equation}
f_q^{A}(x,Q_0^2) = \int_x^A {dy\over y} W_q (y,A) f_q^{p} \left({x \over y},Q_0^2\right),
\end{equation}
in which $W_i (y,A)$ contains the information about the nuclear
effects. For instance, if nuclear effects are disconsidered, $W_q
(y,A) = A \delta (1 - y)$.

\begin{figure}[ht]
\includegraphics[scale=0.45]{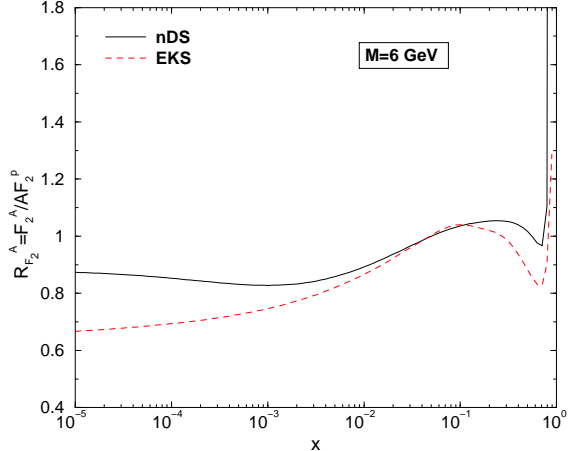}
\caption{\label{cap:Comparison-between-EKS}Comparison between EKS and nDS
parametrizations for the ratio $F_{2}^{A}/F_{2}^{p}$ (A=Au)
for a dilepton mass (scale) $M=6$ GeV as a function of the Bjorken $x$ of the nucleus.}
\end{figure}

The nuclear effects are verified by comparison of the nuclear
structure function with the proton (or deuterium) structure
function. In the Fig.\ \ref{cap:Comparison-between-EKS} the prediction
for the ratio $F_{2}^{A}/AF_{2}^{p}$ is presented for both
parametrizations. The figure can be divided in four regions of
Bjorken-$x$ \cite{Armesto}. In the Fermi motion region $x \gtrsim
0.8$, $R_{F_{2}}^{A}$ is greater than 1 and increases with $x$. The
EMC region $0.3 \lesssim x \lesssim 0.8$ is characterized by
$R_{F_{2}}^{A}<1$. The antishadowing region $0.1 \lesssim x \lesssim
0.3$, and the shadowing region $x \lesssim 0.1$ are defined by
$R_{F_{2}}^{A}>1$, and $R_{F_{2}}^{A}<1$, respectively. The Fig.\
\ref{cap:Comparison-between-EKS} shows that the parametrizations
differ most in both ECM and shadowing regions, with EKS
parametrization getting a lower ratio than nDS.  The parametrizations
are considered at the leading-order, since the approach employed here
considers LO diagrams, without higher orders calculations. Indeed, the
dipole cross section should carry some information about higher
orders, but these properties are included in the phenomenological
parametrization of the dipole cross section, in our case, the GBW
parametrization \cite{GBW}.

In the next section, the results for dilepton production will be
explored, considering both parametrizations for the nuclear parton
distribution presented in this section.

\section{Results and Discussions}

The results for transverse momentum and rapidity distributions for the
dilepton production (with mass $M=6$ GeV) at RHIC ($\sqrt{s}=200$ GeV)
and LHC ($\sqrt{s}=8.8$ TeV) energies at backward rapidities will be
discussed.

\begin{figure}[h]
\includegraphics[scale=0.7]{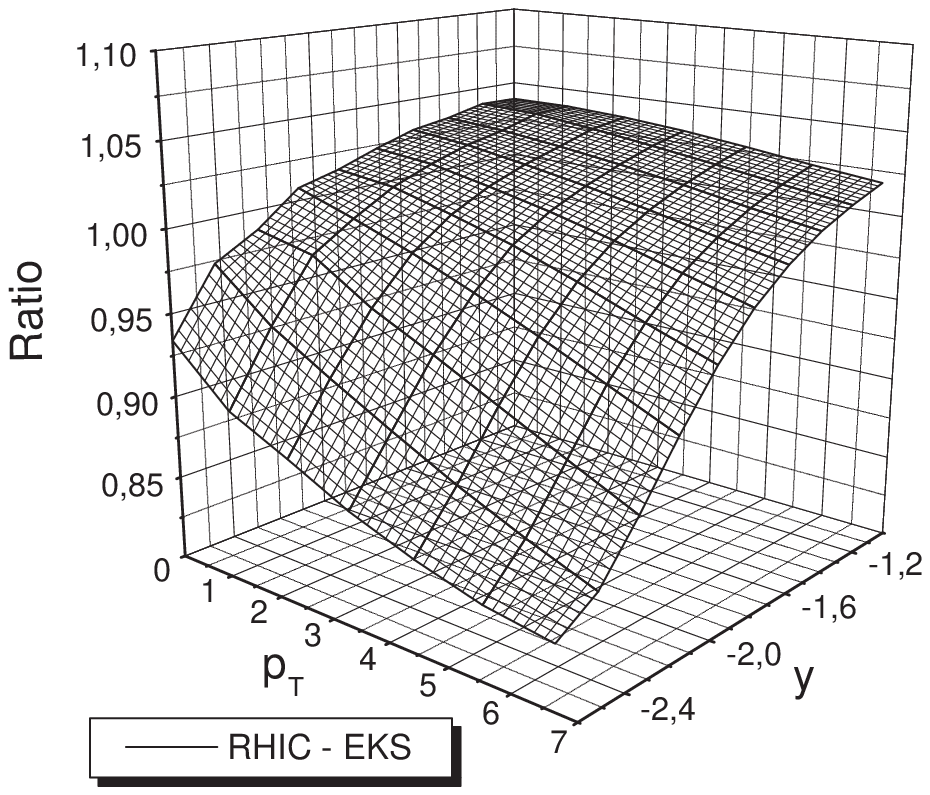}

\vspace{-2.5cm}
\includegraphics[scale=0.7]{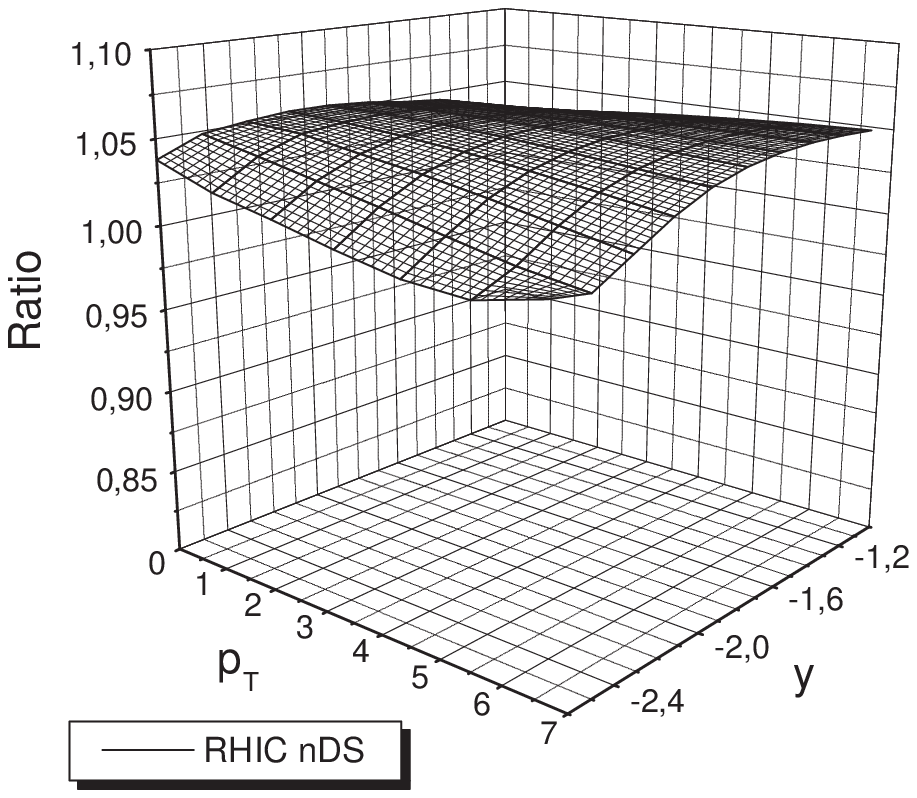}
\vspace{-2cm}~\caption{\label{ratioRHIC}Ratio $R_{pA}$ for RHIC energies considering the
EKS and nDS parametrizations.}
\end{figure}

\begin{figure}[h]
\includegraphics[scale=0.7]{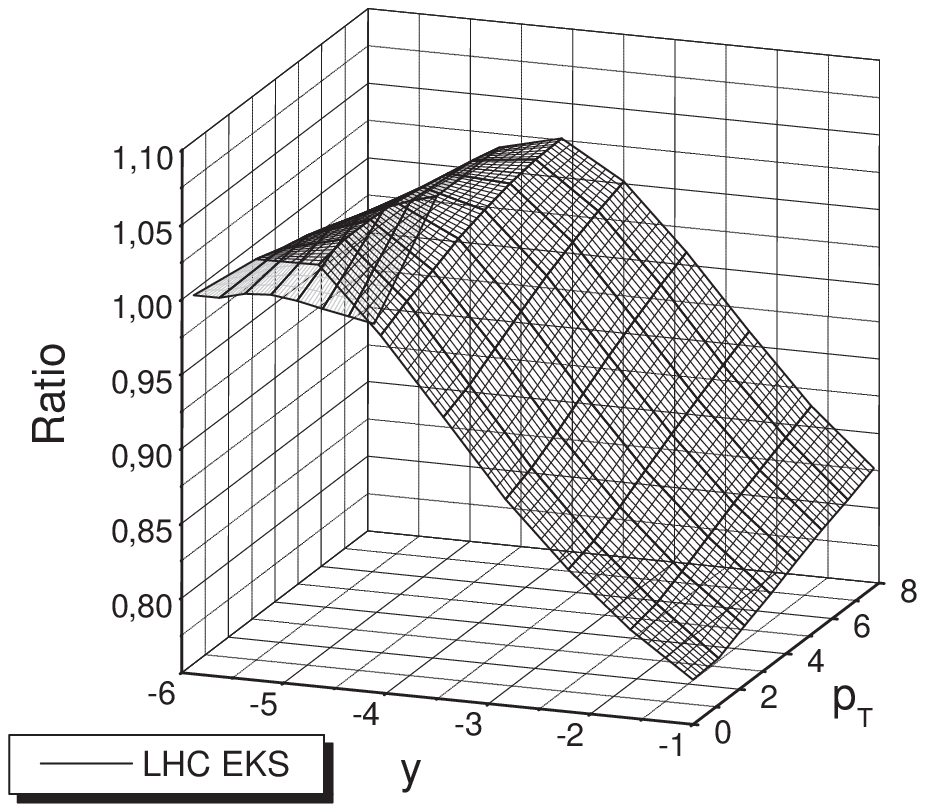}

\vspace{-2.5cm}
\includegraphics[scale=0.7]{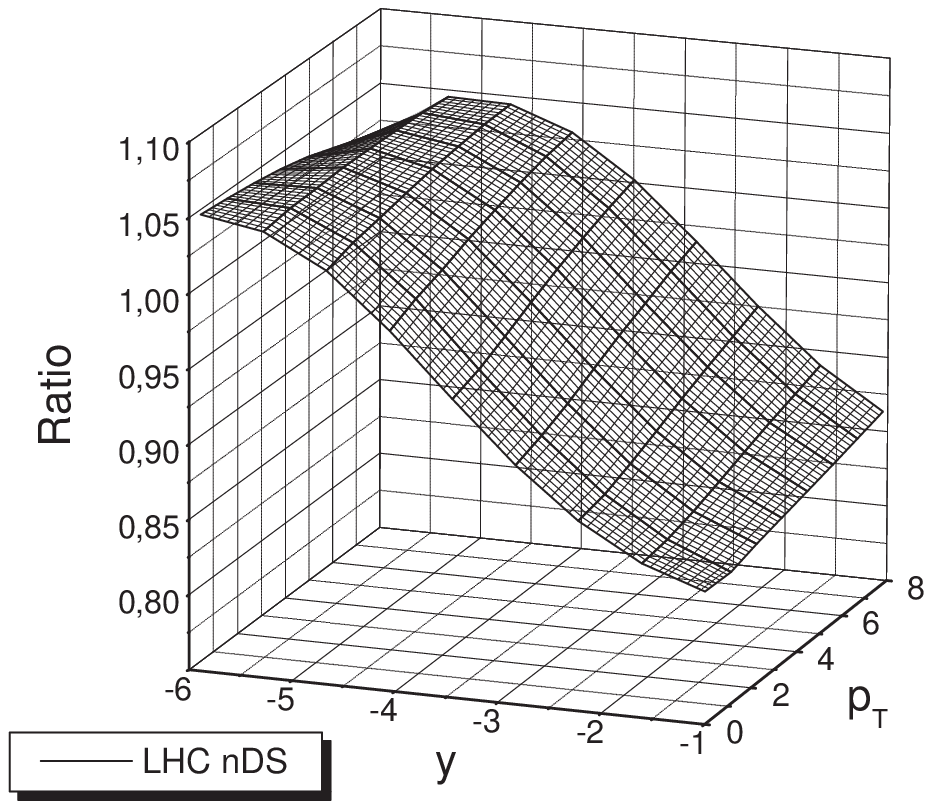}
\vspace{-2cm}
\caption{\label{ratioLHC}Ratio $R_{pA}$ for LHC energies considering the
EKS and nDS parametrizations.}
\end{figure}

The ratio $R_{pA}$ was evaluated at RHIC and LHC energies, and in the
Figs.\ \ref{ratioRHIC} and \ref{ratioLHC} the results are shown in 3D
plots for rapidity and $p_{T}$ spectra, considering the EKS and nDS
parametrizations. The behavior of the ratio $R_{pA}$ reflects the
$x_2$ dependence of the ratio $F_2^A/F_2^p$, presented in the Fig.\
\ref{cap:Comparison-between-EKS}.

At RHIC energies a weak dependence of the ratio on the transverse
momentum is verified, and in general the ratio reaches smaller values
at large $p_{T}$, being this more evident with the EKS
parametrization. Concerning the rapidity spectra, the EKS
parametrization predicts a strong suppression of the ratio at very
backward rapidities and large $p_T$, in comparison with the nDS
parametrization, which predicts an almost flat behavior. 

To explain such results it is necessary to know the range of $x_2$
values in the rapidity and $p_T$ spectra at RHIC. For RHIC energies
considering rapidities from -1 to -2.6, and $p_T$ from 1 to 7, the
$x_2$ range is between 0.08 and 0.5, respectively, meaning that for
more backward rapidities, partons with larger $x_{2}$ are being
probed. The nuclear effects that appear in the $F_{2}^{A}$ at this
$x_{2}$ range are mainly due to EMC effect (reduction of the ratio
$R_{F_2}^A$ as $x_2$ increases, see Fig.\
\ref{cap:Comparison-between-EKS}), which provides the reduction of the
ratio $R_{pA}$ at lower rapidities in the Fig.\ \ref{ratioRHIC}.
Concerning the $p_{T}$ spectra, $x_2$ increases with $p_{T}$, and as
the region probed here is related to the EMC effect, the result is a
reduction of the ratio $R_{pA}$ as $p_T$ increases.  The large
suppression of the ratio $R_{pA}$ of the EKS prediction in comparison
with the nDS in the Fig.\ \ref{ratioRHIC} is a consequence of the
large difference in the $R_{F_2}^A$ predictions of the
parametrizations in the EMC effect region.

At LHC energies (Fig.\ \ref{ratioLHC}) the rapidity spectra present a
peak at intermediated rapidities, which is more pronounced in the EKS
parametrization in comparison with the nDS parametrization. The
$R_{pA}$ transverse momentum dependence presents two distinct
behaviors: for very backward rapidities the ratio decreases as $p_T$
increases and for more central rapidities the ratio $R_{pA}$ increases
with $p_T$.

For LHC energies, rapidities from -1 to -6, and $p_T$ from 1 to 7, the
$x_2$ range is between 0.002 and 0.3, respectively. Here we verify
that not only large $x$ of the nucleus is been probed, but small $x$
too. The $x_{2}$ range probed at LHC provides that shadowing and
antishadowing effects are present.  The peak at intermediate
rapidities is related to the antishadowing effect and the suppression
at more central rapidities is related to the shadowing effect.
The $p_{T}$ spectra is more involved since the ratio $R_{pA}$ presents different behaviors.
For more backward rapidities the ratio is reduced for large $p_{T}$ 
($x_2$ in the antishadowing region, near to EMC region).
For more central rapidities, the ratio increases for large $p_{T}$, since the $x_2$ is in the shadowing region. 

Comparing the predictions from the EKS and nDS parametrizations, in
the Fig. \ref{cap:Comparison-between-EKS} the EKS parametrization
predicts more pronounced antishadowing, which explains the results
found here for the ratio $R_{pA}$ at LHC energies
(Fig. \ref{ratioLHC}).  Our results presented for LHC energies show
that dilepton carries information about the large and small $x$ regime
of the nuclear environment. The distinct nuclear effects are verified
in rapidity and $p_T$ spectra for the dilepton production at backward
rapidities, for RHIC and LHC energies, following the $x$ region
involved.

\begin{figure}[h]
\includegraphics[scale=0.5]{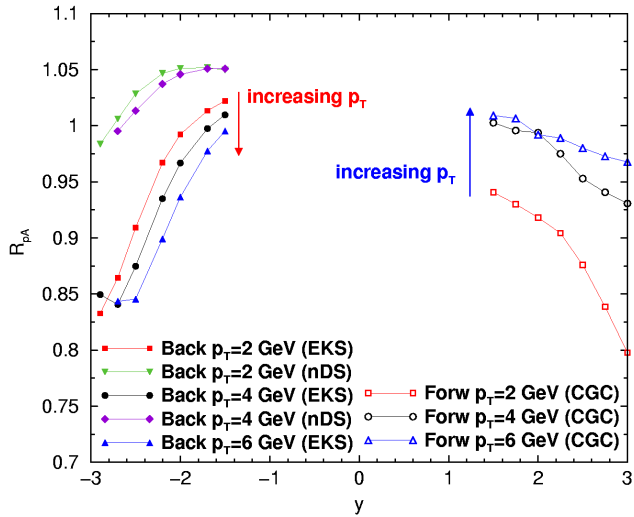}
\caption{\label{ratioRHICBackforw}Comparing the ratio $R_{pA}$ for dileptons at backward and forward rapidities for RHIC energies, considering the
EKS and nDS nuclear parametrizations at backward, and the CGC predictions at forward \cite{PRD70116005}.}
\end{figure}

In order to compare the results presented in this work, with the
previous results from the forward region \cite{PRD70116005}, in the
Fig.~\ref{ratioRHICBackforw} the ratio $R_{pA}$ for RHIC energies is
shown for positive and negative rapidities. The ratio presents
different behaviors concerning the transverse momentum
dependence. While at forward rapidities the saturation phenomena
implies that the ratio $R_{pA}$ increases at large $p_T$, at backward
rapidities the ratio $R_{pA}$ decreases as large $p_T$ is
reached. This distinct behavior is associated with the large $x$
nuclear effects present at backward rapidities.

\section{Conclusions}

The nuclear modification ratio $R_{pA}$ for the dilepton production
was investigated for $p_T$ and backward rapidity spectra in the dipole
picture for RHIC and LHC energies. We have verified a strong
dependence of the nuclear modification ratio with the nuclear
effects. The results presented in this work are explained by the
dependence of the nuclear structure function ratio $R_{F_{2}}^{A}$ on
the Bjorken $x$. The hadron production was investigated at backward
rapidities by PHENIX collaboration at RHIC \cite{backphenix}
presenting the data on the nuclear modification ratio. The results
present an enhancement in the nuclear modification ratio for
$1.5<p_T<4.0$, which still demands some caution, once there are some
data uncertainties and a discrepancy between analysis methods of the
data.  The results presented in this work for dileptons indicate that
the enhacement in the hadron spectra at backward rapidities could be
mainly due to final states, since, the intial state effects
investigated here do not provide such behavior.  Moreover, the
dilepton at backward rapidities is a suitable observable to understand
and quantify the nuclear effects at large and small Bjorken
$x$. Additionally, the transverse momentum dependence of the ratio
$R_{pA}$ is strongly modified at RHIC energies, if forward and
backward rapidities are compared, due to distinct $x$ regions being
probed.

\section*{Acknowledgements}
We thank Boris Kopeliovich for fruitful discussions during his visit
to GFPAE at the occasion of the I LAWHEP and for a careful reading of
this manuscript. This work was partially supported by CNPq, Brazil.

\end{document}